\documentclass[aps,prx,twocolumn,superscriptaddress,nofootinbib]{revtex4-2}
\usepackage{amsmath,amssymb,amsfonts}
\usepackage{graphicx}
\usepackage{hyperref}
\usepackage{xcolor}

\usepackage{amsthm}

\usepackage{fancyhdr}
\usepackage[calcwidth]{titlesec}
\usepackage[font={footnotesize},labelfont={bf}]{caption}
\usepackage{setspace}

\usepackage{extarrows}

\usepackage[most]{tcolorbox}
\newtcolorbox{myt}[2][]{%
  attach boxed title to top center
               = {yshift=-4pt},
  colback      = blue!5!white,
  colframe     = blue!75!black,
  halign       = flush left,
  fonttitle    = \bfseries\sffamily,
  colbacktitle = blue!65!black,
  title        = #2,#1,
  enhanced,
}
\newtcolorbox{myd}[2][]{%
  attach boxed title to top center
               = {yshift=-4pt},
  colback      = violet!5!white,
  colframe     = violet!75!black,
  halign       = flush left,
  fonttitle    = \bfseries\sffamily,
  colbacktitle = violet!65!black,
  title        = #2,#1,
  enhanced,
}
\newtcolorbox{mye}[2][]{%
  attach boxed title to top center
               = {yshift=-4pt},
  colback      = purple!5!white,
  colframe     = purple!75!black,
  halign       = flush left,
  fonttitle    = \bfseries\sffamily,
  colbacktitle = purple!65!black,
  title        = #2,#1,
  enhanced,
}

\newtcolorbox{myg}[2][]{%
  attach boxed title to top center
               = {yshift=-4pt},
  colback      = green!5!white,
  colframe     = green!75!black,
  halign       = flush left,
  fonttitle    = \bfseries\sffamily,
  colbacktitle = green!65!black,
  title        = #2,#1,
  enhanced,
}

\providecommand{\U}[1]{\protect\rule{.1in}{.1in}}

\usepackage{dsfont}

\usepackage{mathtools}

\def\distill{{\rm Distill}}

\def\1{\mathbf{1}}

\def\prob{{\rm Prob}}

\def\pos{{\rm Pos}}

\def\cptp{{\rm CPTP}}

\newcommand{\epm}{\end{pmatrix}}
\newcommand{\bpm}{\begin{pmatrix}}

\renewcommand{\log}{{\operatorname{log}}}

\newcommand{\ebm}{\end{bmatrix}}
\newcommand{\bbm}{\begin{bmatrix}}

\def\bmyd{\begin{myd}{}
\begin{definition}}
\def\emyd{\end{definition}\end{myd}}

\def\bmyl{\begin{myg}{}
\begin{lemma}}
\def\emyl{\end{lemma}\end{myg}}

\def\bmyt{\begin{myt}{}
\begin{theorem}}
\def\emyt{\end{theorem}\end{myt}}

\def\bmyc{\begin{myg}{}
\begin{corollary}}
\def\emyc{\end{corollary}\end{myg}}

\def\>{\rangle}
\def\<{\langle}

\def\mE{\mathcal{E}}

\def\mF{\mathcal{F}}
\def\mN{\mathcal{N}}

\newcommand{\supp}{\operatorname{supp}}

\renewcommand{\ge}{\geqslant}
\renewcommand{\le}{\leqslant}
\renewcommand{\geq}{\geqslant}
\renewcommand{\leq}{\leqslant}

\newcommand{\ben}{\begin{enumerate}}
\newcommand{\een}{\end{enumerate}}

\theoremstyle{definition}
\newtheorem{theorem}{Theorem}
\newtheorem{lemma}[theorem]{Lemma}
\newtheorem{corollary}[theorem]{Corollary}

\newtheorem{definition}[theorem]{Definition}
\newtheorem{remark}[theorem]{Remark}

\usepackage{array}
\usepackage{amsmath}

\newcommand{\bea}{\begin{eqnarray}}
\newcommand{\eea}{\end{eqnarray}}
\newcommand{\be}{\begin{equation}}
\newcommand{\ee}{\end{equation}}
\newcommand{\ba}{\begin{equation}\begin{aligned}}
\newcommand{\ea}{\end{aligned}\end{equation}}

\newcommand{\bee}{\begin{enumerate}}
\newcommand{\eee}{\end{enumerate}}

\def\be{\begin{equation}}
\def\ee{\end{equation}}

\newcommand{\md}{\mathfrak{D}}

\newcommand{\lr}{\rangle\langle}

\newcommand{\tr}{{\rm Tr}}

\newcommand{\eps}{\varepsilon}

\newcommand{\mbb}[1]{\mathbb{#1}}

\newcommand{\eqdef}{\coloneqq}

\def\p{\mathbf{p}}
\def\q{\mathbf{q}}

\def\0{\mathbf{0}}

\def\tD{\widetilde{D}}

\def\tQ{\widetilde{Q}}

\def\d{{\mathrm{d}}}

\begin{document}

\title{Optimal Trace Inequalities for Single-Shot Quantum Information}
\author{Gilad Gour}
\affiliation{Faculty of Mathematics, Technion--Israel Institute of Technology, Haifa 3200003, Israel}

\begin{abstract}
Single-shot quantum information theory is governed not only by entropy exponents, but also by the finite-resource constants that multiply them. These constants directly affect the quantitative performance of decoupling, covering, convex-splitting, position-based decoding, and one-shot communication protocols, yet they are often inherited from nonoptimal scalar estimates or from classical-to-quantum lifting arguments that introduce additional losses. In this work we show that the operator layer-cake representation provides a mechanism for lifting sharp scalar inequalities to the noncommutative setting without loss. Using an iterative Riemann--Stieltjes integration-by-parts method, we derive sharp quantum trace inequalities that tighten several standard single-shot bounds. For a logarithmic trace inequality recently introduced by Cheng \emph{et al.}\ and subsequently used in quantum covering and decoupling problems, we determine the exact optimal prefactor, replacing the previously known constant by a smaller Lambert-$W$ constant and proving universal optimality for positive operators. We also completely characterize the threshold behavior that appears under normalization to quantum states. In addition, we establish optimal two-sided collision-divergence inequalities, which lead to improved position-based decoding and single-shot classical communication bounds. These results show that several finite-resource bounds in single-shot quantum information can be tightened, and that within the layer-cake R\'enyi-divergence framework the resulting constants are genuine optimality barriers rather than artifacts of the proof.
\end{abstract}

\maketitle

\section{Introduction}

Single-shot quantum information theory asks what can be achieved with finite resources: a finite number of channel uses, a finite blocklength, a prescribed error tolerance, and no appeal to asymptotic typicality. This regime is not merely a refinement of asymptotic information theory. It is the natural setting for many operational primitives, including decoupling~\cite{DFR2014,HHWY2008}, convex-splitting~\cite{AJ2019,AJ2022}, covering and soft-covering lemmas~\cite{Cuff2013,GPC2016}, quantum state merging~\cite{HOW2005,ADHW2009}, entanglement distillation~\cite{BBPS1996,DW2005}, and communication over quantum channels~\cite{Devetak2005,HHWY2008}. In these problems, R\'enyi-type information quantities often determine the exponential behavior, but the constants multiplying them determine the quantitative finite-resource performance. Improving such constants can therefore change the number of random codewords needed, the amount of correlation suppression obtained, the achievable message size, or the error guarantees in one-shot protocols.

A central reason these constants are difficult to optimize is that many one-shot quantum information bounds ultimately rely on trace inequalities for noncommuting positive operators. In the classical case, such inequalities often follow from elementary scalar estimates, for example inequalities involving $\log(1+r)$ or $r/(1+r)$. In the quantum case, however, lifting scalar estimates to operator inequalities is subtle. Classical reductions, pinching arguments, and measured-divergence methods may introduce additional losses, while a direct operator-level proof must control threshold projections of noncommuting pencils of the form $\rho-r\sigma$. Thus, even when the relevant exponent is known, the sharp finite-resource prefactor may remain hidden.

Recent progress has shown that the operator layer-cake representation provides a powerful framework for this problem~\cite{CL2025,LHC2025}. This representation expresses several quantum divergences through integrals over projections $\{\rho>r\sigma\}$, and is closely related to Frenkel's integral formula for quantum relative entropy~\cite{Frenkel2023} and its variants~\cite{HT2024}, as clarified in~\cite{CGLL2025}. In particular, Cheng \emph{et al.}~\cite{CGH+2025} used this framework to prove a logarithmic trace inequality of the form
\be
\tr\!\left[\rho\bigl(\log(\rho+\sigma)-\log(\sigma)\bigr)\right]
\le {c_s\over s}\,\tQ_{1+s}(\rho\|\sigma),
\label{eq}
\ee
where $c_s \eqdef s^s(1-s)^{1-s}$ and $\tQ_\alpha \eqdef 2^{(\alpha-1)\tD_\alpha}$ is defined in terms of the sandwiched R\'enyi divergence $\tD_\alpha$. This inequality has become an important input in quantum covering problems and decoupling-type estimates~\cite{CGH+2025,BCY2026}. However, the prefactor $c_s/s$ is not the best possible constant suggested by the logarithmic scalar inequality itself. Since this prefactor propagates directly into finite-resource bounds, it is natural to ask whether it can be sharpened at the fully quantum level, and whether the resulting improvement is optimal.

The first main result of this work answers this question. We prove the sharp logarithmic trace inequality
\be\label{in2}
Q(\rho\|\sigma)\le G_s Q_{1+s}(\rho\|\sigma),
\ee
where
\be
Q(\rho\|\sigma)
\eqdef
\tr\!\left[\rho\bigl(\log(\rho+\sigma)-\log(\sigma)\bigr)\right],
\ee
and $Q_{1+s}$ is the layer-cake R\'enyi divergence~\cite{LHC2025}. The constant $G_s$ is the optimal scalar constant in
\be
\log(1+r)\le G_s r^s,\qquad r\ge0,
\ee
and has a closed-form expression in terms of the Lambert $W$ function. Its improvement over the previous prefactor $c_s/s$ is illustrated in Fig.~\ref{ratio}. Since $Q_{1+s}\le \tQ_{1+s}$, the same constant also improves the corresponding sandwiched-R\'enyi bound. In the limit $s\to0$, this replaces the prefactor $c_s/s$ by a constant smaller by an asymptotic factor of $e$.

\begin{figure}[t]
\includegraphics[width=0.9\columnwidth]{ratio}
\caption{The ratio $R(s)=\frac{c_s}{sG_s}$ as a function of $s$. It ranges from $1$ at $s=1$ to $e$ as $s\to 0$.}
\label{ratio}
\end{figure}

The sharpness statement is equally important. We prove that the constant $G_s$ in~\eqref{in2} is optimal for positive operators: equality is attained by explicit commuting positive operators. Consequently, no smaller constant can make~\eqref{in2} hold uniformly over all finite-dimensional positive operators. The same optimal constant remains valid if $Q_{1+s}$ is replaced by the larger sandwiched quantity $\tQ_{1+s}$, since the equality example is commuting. Hence the improvement is not only a better proof estimate; it identifies the exact endpoint of this class of R\'enyi-divergence bounds.

We then analyze the operationally important case of normalized quantum states. This restriction is not automatic, because the examples proving optimality for arbitrary positive operators need not satisfy $\tr[\rho]=\tr[\sigma]=1$. We show that normalization produces a sharp threshold phenomenon. If
\be
s\le {1\over 2\log(2)},
\ee
then the same constant $G_s$ remains optimal for density matrices. Above this threshold, the optimal constant for normalized states drops to $\log(2)$. Thus the normalized quantum problem is completely characterized.

The second set of results concerns collision-type trace inequalities. We prove that for every $s\in(0,1)$,
\be
Q_2(\rho\|\rho+\sigma)
\ge \tr[\rho]-c_s Q_s(\rho\|\sigma),
\ee
and
\be
Q_2(\rho\|\rho+\sigma)
\le c_s Q_{1+s}(\rho\|\sigma),
\ee
where again $c_s=s^s(1-s)^{1-s}$. The constant $c_s$ is optimal in both inequalities: the same commuting examples attain equality, and no smaller constant works uniformly for all positive operators. Although $c_s$ is not optimal for the logarithmic inequality, it is the sharp constant for these collision-divergence inequalities. In addition, the upper collision bound implies the earlier logarithmic estimate~\eqref{eq}, showing that it is a stronger structural statement. For normalized states, the two collision inequalities behave differently. For the lower collision bound, the same constant $c_s$ remains optimal for all $s\in(0,1)$. By contrast, the upper collision bound exhibits a threshold: the sharp constant is $c_s$ for $s\le\tfrac12$, while for $s>\tfrac12$ it drops to $\tfrac12$. Thus normalization affects the two collision inequalities in distinct ways, both of which can be characterized exactly.

The proofs are based on an iterative Riemann--Stieltjes integration-by-parts method within the operator layer-cake framework~\cite{CL2025,LHC2025}. The method begins with a sharp scalar inequality for a function of the threshold parameter $r$, integrates it against the positive measure generated by the monotone function
\be
r\mapsto \tr[\rho\{\rho>r\sigma\}],
\ee
and then integrates by parts again to recover the relevant layer-cake R\'enyi divergence. The underlying Riemann--Stieltjes viewpoint is already present in~\cite{LHC2025}; here we use it iteratively to lift optimal scalar inequalities to sharp noncommutative trace inequalities. In contrast to classical-to-quantum lifting methods based on pinching, this approach works directly at the operator level and preserves the optimal constant.

The logarithmic inequality can also be combined with classical-to-quantum lifting arguments, leading to variants formulated in terms of measured R\'enyi divergences. Since measured R\'enyi divergences can be smaller than the noncommutative divergences appearing in these bounds, such formulations may improve the divergence term. However, the known lifting procedures introduce pinching-dependent prefactors, as in the approach of Li and Yao~\cite{LY2024}. It is therefore not clear in general whether this trade-off improves the final finite-resource bound.

Finally, we apply the collision inequalities to single-shot communication primitives. First, we obtain an enhanced position-based decoding lemma in which the factor $c_s<1$ improves the number of positions needed for a prescribed decoding error. The result applies not only to the standard product-form position-based construction, but also to a broader class of pairwise index-symmetric states. Second, we apply the same inequality to classical communication over a quantum channel, formulated as the distillation of a noiseless classical channel under local superchannels. This yields a tightened one-shot lower bound on the distillable communication in terms of a R\'enyi mutual information, with the improved finite-resource correction coming directly from the sharp constant $c_\alpha$. In particular, the new positive contribution $\log(1/c_\alpha)=h(\alpha)$ can offset the usual finite-error penalty, and in some error regimes makes the total correction term positive.

Taken together, these results show that several standard single-shot quantum information bounds can be tightened at the level of their finite-resource constants. They also show that, in the regimes treated here, the sharp constants obtained from the corresponding classical scalar inequalities can be achieved in the fully quantum setting. Thus the constants are not artifacts of the proof, but optimal barriers within the layer-cake and R\'enyi-divergence framework, both for arbitrary positive operators and for normalized quantum states. This provides sharper operational estimates and clarifies which parts of current one-shot techniques can, and cannot, be further improved.

\section{Preliminaries}

Throughout the paper, all Hilbert spaces are finite dimensional. We denote by
$\pos(A)$ the set of positive semidefinite operators acting on $A$, and by
$\md(A)\subset\pos(A)$ the set of density matrices. The set $\cptp(A\to B)$
denotes the set of completely positive trace-preserving maps from operators on
$A$ to operators on $B$. For a Hermitian operator $X$, we denote by $\{X>0\}$
the spectral projection onto the positive eigenspace of $X$. In particular, for
$\rho,\sigma\in\pos(A)$ and $r\ge0$, we write
\be
\{\rho>r\sigma\}\eqdef \{\rho-r\sigma>0\}.
\ee

We use the layer-cake R\'enyi quantities introduced in~\cite{CL2025,LHC2025}. These quantities are expressed through threshold projections of the form $\{\rho>r\sigma\}$ and are related to Frenkel's integral representation of quantum relative entropy~\cite{Frenkel2023,HT2024,CGLL2025}. In this paper, $Q_\alpha$ denotes the exponential, rather than logarithmic, R\'enyi-type quantity.
For $\alpha>0$, the layer-cake R\'enyi quantity is defined by~\cite{LHC2025} as
\be
Q_\alpha(\rho\|\sigma)
\eqdef \alpha\int_0^\infty r^{\alpha-1}
\tr\big[\sigma\{\rho>r\sigma\}\big]\,dr\, .
\label{lc1}
\ee
Moreover, for $\alpha>1$ it can also be expressed as
\be
Q_\alpha(\rho\|\sigma)= (\alpha-1)\int_0^\infty r^{\alpha-2}
\tr\big[\rho\{\rho>r\sigma\}\big]\,dr\, .
\label{lc2}
\ee
Whenever expressions involving $\log(\sigma)$ or $Q_\alpha(\rho\|\sigma)$ with $\alpha\ge1$ appear, we implicitly assume $\supp(\rho)\subseteq\supp(\sigma)$.

In the commuting case, both formulas reduce to the classical quantity
$\sum_x p_x^\alpha q_x^{1-\alpha}$. For $\alpha>1$, we will also use the
sandwiched quantity
\be
\tQ_\alpha(\rho\|\sigma)\eqdef 2^{(\alpha-1)\tD_\alpha(\rho\|\sigma)}=\tr\left[\left(\sigma^{\frac{1-\alpha}{2\alpha}}\rho\sigma^{\frac{1-\alpha}{2\alpha}}\right)^\alpha\right],
\ee
where $\tD_\alpha$ is the sandwiched R\'enyi divergence. By the Araki-Lieb-Thirring inequality~\cite{LT1976,Araki1990}, one has~\cite{BHT2025,LHC2025}:
\be\label{qa}
Q_\alpha(\rho\|\sigma)\le \tQ_\alpha(\rho\|\sigma)\,,
\qquad \alpha>1\, .
\ee

The case $\alpha=2$ plays a special role. In this case,
\be\label{q2}
Q_2(\rho\|\sigma)
=
\int_0^\infty
\tr\big[\rho\{\rho>r\sigma\}\big]\,dr ,
\ee
and it can also be expressed as
\be
Q_2(\rho\|\sigma)
\eqdef
\frac{d}{dr}\tr\left[\rho\log(\sigma+r\rho)\right]\Big|_{r=0}.
\label{eq:Q2}
\ee
This expression is the quadratic form associated with the
Bogoliubov--Kubo--Mori metric.

Motivated by the logarithmic trace quantity introduced in~\cite{CGH+2025}, we
define
\be
Q(\rho\|\sigma)
\eqdef
\tr\left[\rho\big(\log(\rho+\sigma)-\log\sigma\big)\right].
\label{eq:Qdef}
\ee
In particular, $Q(\rho\|\rho)=\tr[\rho]\log(2)$, and therefore
$Q(\rho\|\rho)=\log(2)$ for normalized states.

A useful feature of $Q$ is that it satisfies the data-processing inequality.
This is not immediate from its definition, since it can be written as the
difference
\be
Q(\rho\|\sigma)
=
D(\rho\|\sigma)-D(\rho\|\rho+\sigma),
\label{eq:Qdiff}
\ee
and differences of monotone quantities are not generally monotone. Instead, the
DPI follows from the integral representation~\cite{CGH+2025}
\be
Q(\rho\|\sigma)
=
\int_0^1 Q_2(\rho\|\sigma+t\rho)\,dt .
\label{eq:Qint}
\ee
Since $Q_2$ satisfies the DPI, the same property follows for $Q$ by integration.

The logarithmic quantity $Q$ also admits the layer-cake representation~\cite{CGH+2025}
\be
Q(\rho\|\sigma)
=
\int_{0}^\infty
\frac{dr}{1+r}\,
\tr\big[\rho\{\rho>r\sigma\}\big].
\label{q}
\ee
This formula is the starting point for the logarithmic trace inequalities proved
below. In the classical case, where $\rho$ and $\sigma$ commute, $Q$ reduces to
the $f$-divergence associated with $f(r)=r\log(1+r)$. In the noncommutative
setting, however, it does not coincide in general with the standard Petz,
minimal, or maximal quantum extensions of this classical $f$-divergence.

We next recall the Riemann--Stieltjes formulation of the layer-cake
representation, which will be used throughout the proofs. Given
$\rho,\sigma\in\pos(A)$, for every $r>0$ define
\begin{align}
\mu(r)&\eqdef -\tr\big[\rho\{\rho>r\sigma\}\big]\,,\label{mur}
\\
\nu(r)&\eqdef -\tr\big[\sigma\{\rho>r\sigma\}\big] \,.\label{nur}
\end{align}
The functions
\be
r\mapsto \tr\big[\rho\{\rho>r\sigma\}\big],
\qquad
r\mapsto \tr\big[\sigma\{\rho>r\sigma\}\big]
\ee
are nonincreasing and right-continuous. Hence $\mu$ and $\nu$ are
nondecreasing functions of bounded variation on finite intervals, and their
distributional derivatives define positive Riemann--Stieltjes measures,
denoted by $\mathrm{d}\mu$ and $\mathrm{d}\nu$.

In this notation, for every $\alpha>0$ and $\alpha\ne 1$, $Q_\alpha$ can be expressed as~\cite{LHC2025}
\be\label{use}
Q_{\alpha}(\rho\|\sigma)
=
\int_0^\infty r^{\alpha}\,\mathrm{d}\nu(r)=\int_0^\infty r^{\alpha-1}\,\mathrm{d}\mu(r)\,.
\ee
The relation between the two Stieltjes measures follows from Lemma~6.3 of~\cite{LHC2025}. Specifically, for every measurable function $g(r)$ for which the integrals
are well defined,
\be\label{munu}
\int_0^\infty g(r)\,\mathrm{d}\mu(r)
=
\int_0^\infty r\,g(r)\,\mathrm{d}\nu(r)\,.
\ee
In particular, this relation yields
\be
\int_0^\infty r\,\mathrm{d}\nu(r)
=
\tr[\rho]\,.
\ee

\section{Optimal Logarithmic Trace Inequalities}

Let $W$ denote the Lambert $W$ function, defined as the multivalued inverse of
$w\mapsto we^w$. On the interval $(-e^{-1},0)$ this inverse has two real
branches: the principal branch $W_0$, taking values in $(-1,0)$, and the lower
branch $W_{-1}$, taking values in $(-\infty,-1)$. In what follows we use the
principal branch and define, for $s\in(0,1)$,
\be\label{r}
r \eqdef -\frac{1}{sW_0\!\left(-\frac{1}{s}e^{-1/s}\right)} - 1\,,
\ee
and
\be\label{eq:Gs}
G_s \eqdef \frac{r^{1-s}}{s(1+r)}\,.
\ee
Then:
\bmyt\label{thm:1}
For every $\rho,\sigma\in\pos(A)$,
\be\label{eq:main-Gs-bound}
Q(\rho\|\sigma)\leq G_s\,Q_{1+s}(\rho\|\sigma).
\ee
Moreover, the constant $G_s$ is optimal, with equality attainable.
\emyt
\noindent\textbf{Remarks.}
\ben
\item Since $Q_{1+s}(\rho\|\sigma)\le \tQ_{1+s}(\rho\|\sigma)$, Eq.~\eqref{eq:main-Gs-bound} immediately implies
\be\label{qta}
Q(\rho\|\sigma)\leq G_s\tQ_{1+s}(\rho\|\sigma)\;.
\ee
Moreover, the same constant $G_s$ is optimal also in~\eqref{qta}. Indeed, the equality example used below is a strictly positive commuting pair, and for commuting operators one has
$Q_{1+s}(\rho\|\sigma)=\tQ_{1+s}(\rho\|\sigma)$.
\item The constant $G_s$ is strictly smaller than the prefactor $c_s/s$ of
\cite{CGH+2025}. Indeed, from~\eqref{eq:Gs},
\be
G_s \leq \frac1s \sup_{r>0}\frac{r^{1-s}}{1+r}
= \frac{c_s}{s},
\ee
where the supremum is attained at $r=(1-s)/s$. The ratio $R(s)=c_s/(sG_s)$ is shown in Fig.~\ref{ratio}, and satisfies
$R(s)\to e$ as $s\to0^+$.
\een

\begin{proof}
The smallest constant $G_s$ that satisfies
\be\label{gs}
\log(1+r)\leq G_s r^s\qquad \forall\,r>0\,,
\ee
is given by
\be
G_s=\sup_{r>0}\frac{\log(1+r)}{r^s}\,.
\label{eq:Gs-def}
\ee
The function
$r\mapsto r^{-s}\log(1+r)$ vanishes at both endpoints $r=0$ and
$r=\infty$, and its critical points satisfy
\be
r=s(1+r)\log(1+r)\,.
\label{eq:crit}
\ee
This equation has a unique solution $r>0$, which is therefore the maximizer.
Setting
$
t\eqdef -\frac{1}{s(1+r)}
$,
one finds that~\eqref{eq:crit} is equivalent to
\be
t e^t = -\frac{1}{s}e^{-1/s}\,.
\ee
Thus the positive maximizer is obtained from the principal real branch of the
Lambert $W$ function:
\be
t=W_0\!\left(-\frac1s e^{-1/s}\right),
\ee
or equivalently
\be
r=-\frac{1}{sW_0\!\left(-\frac1s e^{-1/s}\right)}-1\,.
\ee
Using the critical-point equation~\eqref{eq:crit}, we obtain~\eqref{eq:Gs}.

We now extend the scalar inequality~\eqref{gs} to establish~\eqref{eq:main-Gs-bound}. Using the layer-cake representation~\eqref{q}, let $\mu(r)$ be as in~\eqref{mur}, $g(r)\eqdef\log(1+r)$, and set
\be\label{rbig}
R\eqdef \big\|\sigma^{-1/2}\rho\,\sigma^{-1/2}\big\|_\infty\,.
\ee
Then $\mu(r)=0$ for $r\ge R$ so that~\eqref{q} can be written as
\be
Q(\rho\|\sigma)=-\int_0^R g'(r)\mu(r)\,dr\,.
\ee
Since $g$ is absolutely continuous, Stieltjes integration by parts yields
\ba\label{40}
Q(\rho\|\sigma)
&=-\mu(R)g(R)+\mu(0)g(0)
+\int_0^R g(r)\,\mathrm{d}\mu(r)\\
&=
\int_0^R g(r)\,\mathrm{d}\mu(r)\,,
\ea
where we used $\mu(R)=0$ and $g(0)=0$.
Applying the pointwise bound $g(r)\le G_s r^s$ and using positivity of $\mu$,
\ba
Q(\rho\|\sigma)
\le
G_s\int_0^R r^s\,\mathrm{d}\mu(r)&=
G_s\int_{0}^\infty r^s\,\mathrm{d}\mu(r)\\
&=G_sQ_{1+s}(\rho\|\sigma)\,.
\label{middle}
\ea
where the last equality follows from~\eqref{use}.

It remains to prove the optimality of the constant $G_s$. Consider the commuting pair
\be\label{rhosig}
\rho=\frac{1}{k}\Pi_k\,,
\qquad
\sigma=\frac{\lambda}{d}I_d\,,
\ee
where $\Pi_k$ is a rank-$k$ projection with $k\le d$, and $\lambda>0$. For this choice, both quantities depend only on the single parameter
$
r\eqdef d/(\lambda k)
$.
A direct computation yields
\be\label{qr}
Q(\rho\|\sigma)=\log(1+r)\,,
\qquad
Q_{1+s}(\rho\|\sigma)=r^s\,.
\ee
Since $\lambda>0$ is arbitrary, any $r>0$ can be realized. Choosing $r$ to be the maximizer of
$r\mapsto \log(1+r)/r^s$, we obtain
\be
\frac{Q(\rho\|\sigma)}{Q_{1+s}(\rho\|\sigma)}
=
\frac{\log(1+r)}{r^s}
=
G_s\,.
\ee
Thus the constant $G_s$ cannot be improved, which establishes optimality.
\end{proof}

\begin{remark}
The identity~\eqref{use}, taken from~\cite{LHC2025}, is itself derived by
Riemann--Stieltjes integration by parts. The proof above therefore uses an
integration-by-parts loop: one integration-by-parts step converts the
 layer-cake integral~\eqref{q} into a logarithmic integral against $\mathrm{d}\mu$ (see~\eqref{40}), while
the known identity~\eqref{use} converts the resulting power integral back into
the layer-cake R\'enyi quantity~\eqref{lc2}. This loop is the mechanism that preserves the
sharp scalar constant $G_s$ at the operator level.
\end{remark}

\section{Two Trace Inequalities for the Collision Divergence}

In this section, we present another application of the Riemann--Stieltjes integration-by-parts loop, leading to two new trace inequalities. The result is expressed in terms of the quantum collision divergence.
Classically, for diagonal matrices $\rho=\p$ and $\sigma=\q$, the collision divergence is given by
\be
Q_2(\p\|\q)=\sum_{x\in[d]} \frac{p_x^2}{q_x}\,.
\ee
In the quantum setting, we use the collision layer-cake divergence as given
in~\eqref{q2}.
With this, we establish the following theorem.

Let $s\in(0,1)$ and $c_s\eqdef s^s(1-s)^{1-s}$.  Then, for every $\rho,\sigma\in\pos(A)$:
\bmyt\label{thm:3}
\begin{align}
Q_2(\rho\|\rho+\sigma)&\geq \tr[\rho]-c_s Q_s(\rho\|\sigma)\label{lb}\\
Q_2(\rho\|\rho+\sigma)&\leq c_s Q_{1+s}(\rho\|\sigma)\,.\label{ub}
\end{align}
Moreover, the constant $c_s$ is optimal in both inequalities, with equality attainable.
\emyt

\begin{remark}
The inequality~\eqref{ub} is in fact stronger than~\eqref{eq}. Indeed, using
the integral representation~\eqref{eq:Qint} and the homogeneity relation
$Q_{1+s}(\rho\|t\sigma)=t^{-s}Q_{1+s}(\rho\|\sigma)$ for $t,s>0$, we obtain
\ba
Q(\rho\|\sigma)&=\int_0^1 dt\; Q_2(\rho\|\sigma+t\rho)\\
&=\int_0^1 dt\; \frac1tQ_2\Bigl(\rho\big\|\rho+\frac1t\sigma\Bigr)\\
&\leq c_s\int_0^1 dt\; \frac1t  Q_{1+s}\Big(\rho\big\|\frac1t\sigma\Big)\\
&=c_sQ_{1+s}(\rho\|\sigma)\int_0^1 dt\; t^{s-1}\\
&=\frac{c_s}s  Q_{1+s}(\rho\|\sigma)\,,
\ea
where the inequality follows from~\eqref{ub}.
\end{remark}

\begin{proof}[Proof of the Upper Bound~\eqref{ub}]
We first show that the coefficient $c_s$ satisfies, for all $p,q>0$,
\be\label{inx0}
\frac{p^2}{p+q}\le c_s\, p^{1+s} q^{-s}\,.
\ee
To see this, set $r\eqdef p/q$. Then~\eqref{inx0} is equivalent to
\be\label{rcs}
\frac{r^{1-s}}{1+r}\le c_s\,.
\ee
For $s\in(0,1)$, the function $r\mapsto r^{1-s}/(1+r)$ attains its maximum at $r=(1-s)/s$, with value
$
c_s = s^s(1-s)^{1-s}
$.
Thus~\eqref{inx0} holds for all $p,q>0$.
 
 We now lift this inequality to the quantum setting. First observe that
\ba\label{q3}
 Q_2(\rho\|\rho+\sigma)&=\int_{0}^\infty dr\,\tr\big[\rho\{\rho>r(\rho+\sigma)\}\big]\\
 &=\int_{0}^1 dr\,\tr\big[\rho\{\rho>r(\rho+\sigma)\}\big]\\
 &=\int_{0}^1 dr\,\tr\left[\rho\left\{\rho>\frac r{1-r}\sigma\right\}\right]\\
&=\int_{0}^\infty \frac{dt}{(1+t)^2}\,\tr\big[\rho\{\rho>t\sigma\}\big]\;,
 \ea
 where in the last step we used the change of variables $t\eqdef\frac{r}{1-r}$.

 We now apply the Riemann--Stieltjes integration-by-parts loop. Let $\mu(r)$ be as in~\eqref{mur}, define
\be
g(t)\eqdef \frac{t}{1+t}\,,
\ee
and let $R$ be as in~\eqref{rbig}. Then $\mu(R)=0$ and $g(0)=0$. 
 From~\eqref{q3} we obtain 
 \ba\label{gg}
 Q_2(\rho\|\rho+\sigma)&=-\int_{0}^\infty g'(t)\mu(t)\,dt\\&=-\int_{(0,R]} \mu(t)\,\mathrm{d}g\\
 &=-\mu(R)g(R)+\mu(0)g(0)+\int_{(0,R]} g(t)\,\mathrm{d}\mu(t)\\
&=\int_{0}^{\infty} \frac t{1+t}\,\mathrm{d}\mu(t)\;,
 \ea
 where we used $\mu(R)=g(0)=0$. We now use this formula to obtain both the upper and lower bounds.
 
For the upper bound, using~\eqref{rcs}, we have $\frac{t}{1+t}\le c_s t^s$ for all $t\ge 0$. Since $\mathrm{d}\mu(t)\ge 0$, it follows that
\ba\label{qqq}
 Q_2(\rho\|\rho+\sigma)&\le c_s\int_{0}^\infty t^s\,\mathrm{d}\mu(t)\\
&=c_sQ_{1+s}(\rho\|\sigma)\,,
 \ea
 where the last equality follows from~\eqref{use}.
 This completes the proof of~\eqref{ub}.
\end{proof}

\begin{proof}[Proof of the Lower Bound~\eqref{lb}]
For the lower bound, we start again from~\eqref{gg}:
\be
Q_2(\rho\|\rho+\sigma)
=
\int_0^\infty \frac{t}{1+t}\,\mathrm{d}\mu(t).
\ee
Using~\eqref{rcs}, we have
$\tfrac{1}{1+t}\le c_s t^{s-1}$
for $t>0$. Hence,
\ba
Q_2(\rho\|\rho+\sigma)
&=
\int_0^\infty \left(1-\frac{1}{1+t}\right)\,\mathrm{d}\mu(t)\\
&\ge
\int_0^\infty \left(1-c_s t^{s-1}\right)\,\mathrm{d}\mu(t)\\
&=
\tr[\rho]-c_s Q_s(\rho\|\sigma),
\ea
where the last equality follows from~\eqref{use}.
This proves~\eqref{lb}.
\end{proof}

\begin{proof}[Proof of Optimality.]
Take $\rho$ and $\sigma$ as in~\eqref{rhosig}, and set
$
r\eqdef \tfrac{k\lambda}{d}\,.
$
For this commuting pair, a direct computation gives
\be
Q_s(\rho\|\sigma)=r^{1-s}\,,
\qquad
Q_{1+s}(\rho\|\sigma)=r^{-s}\,,
\ee
and
\be
Q_2(\rho\|\rho+\sigma)=\frac{1}{1+r}\,.
\ee
Since $\tr[\rho]=1$, we have
\be\label{rr}
\frac{\tr[\rho]-Q_2(\rho\|\rho+\sigma)}{Q_s(\rho\|\sigma)}
=
\frac{Q_2(\rho\|\rho+\sigma)}{Q_{1+s}(\rho\|\sigma)}
=
\frac{r^s}{1+r}\,.
\ee
The function $r\mapsto r^s/(1+r)$ is maximized at
\be
r=\frac{s}{1-s},
\ee
and its maximum value is
\be
\frac{r^s}{1+r}=s^s(1-s)^{1-s}=c_s\,.
\ee
Choosing $\lambda$ so that $k\lambda/d=s/(1-s)$ therefore gives equality in both~\eqref{lb} and~\eqref{ub}. Hence the constant $c_s$ cannot be improved in either inequality.
\end{proof}

Recently, another relation involving $Q_2(\rho\|\sigma)$ was established in~\cite{Gour2025a}:
\be\label{rps}
Q_2(\rho\|\rho+\sigma)\ge\tr(\rho-\sigma)_+\;.
\ee
This inequality can be applied to provide a straightforward proof of a related inequality established in~\cite{Cheng2023}:
\be\label{mainineq}
\tr[\rho\land\sigma]\geq\tr\left[\rho\Lambda^{-1/2}\sigma\Lambda^{-1/2}\right] \;,
\ee
where $\Lambda\eqdef\rho+\sigma$ and
\ba
\rho\land \sigma\eqdef\frac12(\rho+\sigma-|\rho-\sigma|)\;.
\ea

We now compare these bounds with known estimates involving $Q_s(\rho\|\sigma)$. In~\cite{ACM+2007} it was shown that, for $\rho,\sigma\ge0$ and $s\in[0,1]$, trace quantities involving $\rho\land\sigma$ and $(\rho-\sigma)_+$ can be bounded in terms of $\tr[\rho^s\sigma^{1-s}]$. These bounds were recently sharpened in~\cite{BHT2025} by replacing $\tr[\rho^s\sigma^{1-s}]$ with the smaller layer-cake divergence $Q_s(\rho\|\sigma)$. In particular, one obtains
\be\label{f1}
\tr[\rho \land \sigma] \leq Q_s(\rho\|\sigma)\;,
\ee
or equivalently
\be\label{f2}
\tr(\rho-\sigma)_+\geq \tr[\rho]-Q_s(\rho\|\sigma)\;.
\ee
Combining~\eqref{f2} with~\eqref{rps} gives
\be
Q_2(\rho\|\rho+\sigma)\ge\tr[\rho]-Q_s(\rho\|\sigma)\,.
\ee
Since $c_s<1$ for $s\in(0,1)$, this last bound is weaker than~\eqref{lb}.

On the other hand, combining~\eqref{f1} with~\eqref{mainineq} gives
\be
\tr\left[\rho\Lambda^{-1/2}\sigma\Lambda^{-1/2}\right]\leq Q_s(\rho\|\sigma)\;.
\ee
This inequality is a useful tool in position-based decoding lemmas~\cite{Cheng2023}. 
 By applying the sharper lower bound~\eqref{lb} directly, we obtain the following strengthened version of the position-based decoding estimate.

\bmyc\label{cor1}
Let $\rho,\sigma\in\pos(A)$ and $\Lambda=\rho+\sigma$. Then, for every $s\in(0,1)$
\be
\tr\left[\rho\Lambda^{-1/2}\sigma\Lambda^{-1/2}\right]\leq c_s Q_s(\rho\|\sigma)\;.
\ee
\emyc

\begin{proof}
By definition,
\ba
\tr\left[\rho\Lambda^{-1/2}\sigma\Lambda^{-1/2}\right]&=\tr\left[\rho\Lambda^{-1/2}(\rho+\sigma-\rho)\Lambda^{-1/2}\right]\\
&=\tr[\rho]-\tr\left[\rho\Lambda^{-1/2}\rho\Lambda^{-1/2}\right]\\
&=\tr[\rho]-\tQ_2(\rho\|\rho+\sigma)\\
&\le \tr[\rho]-Q_2(\rho\|\rho+\sigma)\\
&\le c_s Q_s(\rho\|\sigma)\;,
\ea
where the first inequality uses $Q_2\le \tQ_2$, and the last inequality follows from~\eqref{lb}.
\end{proof}

\section{Optimal Bounds for Normalized States}

So far, the trace inequalities and optimality statements have been formulated for arbitrary positive semidefinite operators $\rho,\sigma\in\pos(A)$. The examples establishing optimality in this general setting do not necessarily satisfy the normalization constraints $\tr[\rho]=\tr[\sigma]=1$; in particular, the operator $\sigma$ in~\eqref{rhosig} need not be a density matrix. It is therefore not automatic that the same optimal constants remain valid after restricting to normalized states, which is the setting most directly relevant to quantum information applications. In this section, we resolve the normalized problem. We show that normalization affects different inequalities in different ways: for some of them the optimal constants remain unchanged, while for others normalization produces sharp threshold phenomena, with the constants dropping to smaller values in complementary ranges of $s$ that we determine exactly.

Let $s_0\eqdef 1/(2\log(2))$, and let $G_s$ be defined as in~\eqref{eq:Gs}. Then:
\bmyt
For every $s\in(0,1)$,
\be\label{mainnorm}
\sup_{\rho,\sigma\in\md(A)}
\frac{Q(\rho\|\sigma)}{Q_{1+s}(\rho\|\sigma)}=\begin{cases}
G_s & \text{if }s\leq s_0\\
 \log(2)&\text{if }s>s_0
\end{cases}\,.
\ee
\emyt

\begin{remark}
The supremum in~\eqref{mainnorm} is taken jointly over all finite dimensions of the system $A$ and over all $\rho,\sigma\in\md(A)$. Moreover, at the threshold $s=s_0$ one has $G_{s_0}=\log(2)$, so the two branches in~\eqref{mainnorm} agree.
\end{remark}

\begin{proof}
We start with the case $s\le s_0$. The upper bound
\be
\frac{Q(\rho\|\sigma)}{Q_{1+s}(\rho\|\sigma)}\le G_s
\ee
follows immediately from the corresponding bound for arbitrary positive
semidefinite operators, since $\md(A)\subset\pos(A)$. It remains to show that
the same constant $G_s$ cannot be improved under the normalization constraint.

Consider the example from~\eqref{rhosig} with $\lambda=1$. Then
$\sigma=I_d/d$ is the maximally mixed state, and the parameter in~\eqref{qr} is
$
r=d/k\geq1.
$
By varying $d$ and $k$, the attainable values of $r=d/k$ are dense in
$[1,\infty)$. Hence, using~\eqref{qr} and the continuity of
$r\mapsto \log(1+r)/r^s$, we obtain
\be\label{lower2}
\sup_{\rho,\sigma\in\md(A)}
\frac{Q(\rho\|\sigma)}{Q_{1+s}(\rho\|\sigma)}
\geq
\sup_{r\geq1}\frac{\log(1+r)}{r^s}\;.
\ee
In Appendix~\ref{A1} we show that, for $s\le s_0$, the maximizer of
$r\mapsto \log(1+r)/r^s$ lies in the region $r\ge1$. Thus, the right-hand side
of~\eqref{lower2} equals $G_s$, and we conclude that the normalization
constraint does not reduce the optimal value in the range $s\le s_0$.

Next we consider the case $s>s_0$. To see that $\log(2)$ is a lower bound, observe that, for any $\rho\in\md(A)$,
\be\label{lower3}
\sup_{\rho,\sigma\in\md(A)}
\frac{Q(\rho\|\sigma)}{Q_{1+s}(\rho\|\sigma)}
\geq
\frac{Q(\rho\|\rho)}{Q_{1+s}(\rho\|\rho)}
=
\log(2)\;.
\ee
To show that $\log(2)$ is also an upper bound, let $\mu(r)$ be as in~\eqref{mur}. Then~\eqref{40} gives
\be\label{qint}
Q(\rho\|\sigma)=\int_{0}^\infty \log(1+r)\,\d\mu(r)\,,
\ee
where $\d\mu(r)$ is supported on $(0,R]$. In Appendix~\ref{A2} we show that, for $s>s_0$, the scalar inequality
\be\label{auxineq}
\frac{\log(1+r)}{\log(2)}
\le r^s
+(s-s_0)(r^{-1}-1)
\ee
holds for all $r>0$. Combining this with~\eqref{qint} yields
\be
\frac{Q(\rho\|\sigma)}{\log(2)}
\le
\int_{0}^\infty
\Big(r^s+(s-s_0)(r^{-1}-1)\Big)\,\d\mu(r)\,.
\ee
From~\eqref{use} we have
\be\label{int1s}
\int_{0}^\infty r^s\,\d\mu(r)
=
Q_{1+s}(\rho\|\sigma)\;.
\ee
Moreover,
\be
\int_{0}^\infty 1\,\d\mu(r)=\tr[\rho]=1\;.
\ee
Applying~\eqref{munu} to $g_\varepsilon(r)=(r+\varepsilon)^{-1}$ gives
\be
\int_{(0,\infty)} \frac{1}{r+\varepsilon}\,\d\mu(r)
=
\int_{(0,\infty)} \frac{r}{r+\varepsilon}\,\d\nu(r).
\ee
By monotone convergence, as $\varepsilon\downarrow0$ this yields
\be
\int_{(0,\infty)} r^{-1}\,\d\mu(r)
=
\int_{(0,\infty)} \d\nu(r)
=
\tr[\sigma\Pi_\rho]\,,
\ee
where  $\Pi_\rho$ is the projection onto the support of $\rho$. Since $\tr[\sigma\Pi_\rho]\le 1$, we conclude that
\be\label{leq}
\int_{0}^\infty (r^{-1}-1)\,\d\mu(r)\le 0\;.
\ee
Hence,
\be
\frac{Q(\rho\|\sigma)}{\log(2)}
\le
Q_{1+s}(\rho\|\sigma)\;.
\ee
This proves that $\log(2)$ is an upper bound for $s>s_0$. Together with~\eqref{lower3}, this completes the proof.
\end{proof}

We now revisit the trace inequalities of Theorem~\ref{thm:3} under the additional assumption that the operators are normalized states. We first show that the lower collision-divergence inequality~\eqref{lb} is unaffected by normalization: its optimal constant remains $c_s$ for all $s\in(0,1)$.

\bmyt
For every $s\in(0,1)$,
\be\label{o1}
\sup_{\rho,\sigma\in\md(A)}
\frac{1-Q_2(\rho\|\rho+\sigma)}{Q_s(\rho\|\sigma)}
=
c_s\,.
\ee
\emyt
\begin{proof}
The upper bound follows immediately from Theorem~\ref{thm:3}. Indeed, since
$\rho,\sigma\in\md(A)$ imply $\tr[\rho]=1$, the lower bound in
Theorem~\ref{thm:3} gives
\be
1-Q_2(\rho\|\rho+\sigma)
\le c_s Q_s(\rho\|\sigma)\,,
\ee
and hence
\be
\sup_{\rho,\sigma\in\md(A)}
\frac{1-Q_2(\rho\|\rho+\sigma)}{Q_s(\rho\|\sigma)}
\le c_s\, .
\ee

It remains to prove the reverse inequality. Let
\be
t_s\eqdef \frac{s}{1-s},
\ee
and define two commuting density matrices by
\be
\rho=\operatorname{Diag}(p,1-p)\,,
\qquad
\sigma=\operatorname{Diag}(q,1-q)\,,
\ee
where
\be
p\eqdef \min\left\{1,t_s^{-1}\right\},
\qquad
q\eqdef \min\{1,t_s\}\,.
\ee
Then $q=t_s p$, and at most one of $1-p$ and $1-q$ is nonzero. Therefore the
only nontrivial contribution to both
$1-Q_2(\rho\|\rho+\sigma)$ and $Q_s(\rho\|\sigma)$ comes from the first
coordinate. A direct computation gives
\be
1-Q_2(\rho\|\rho+\sigma)
=
\frac{pq}{p+q}
=
\frac{pt_s}{1+t_s},
\ee
and
\be
Q_s(\rho\|\sigma)
=
p^s q^{1-s}
=
pt_s^{1-s}\,.
\ee
Hence
\be
\frac{1-Q_2(\rho\|\rho+\sigma)}{Q_s(\rho\|\sigma)}
=
\frac{pt_s/(1+t_s)}{pt_s^{1-s}}
=
\frac{t_s^s}{1+t_s}
=
c_s\, .
\ee
Thus the supremum is at least $c_s$. Combined with the upper bound, this proves
\eqref{o1}.
\end{proof}

Finally, we revisit the trace inequality~\eqref{ub} under the additional assumption that the operators are normalized states. We show that normalization produces a sharp threshold phenomenon for this upper collision-divergence inequality. 

Let $c_s\eqdef s^s(1-s)^{1-s}$. Then:
\bmyt
For every $s\in(0,1)$,
\be\label{o2}
\sup_{\rho,\sigma\in\md(A)}
\frac{Q_2(\rho\|\rho+\sigma)}{Q_{1+s}(\rho\|\sigma)}
=
\begin{cases}
c_s & \text{if } s\le \tfrac12,\\
\tfrac12 & \text{if } s>\tfrac12,
\end{cases}\,.
\ee
\emyt
\begin{proof}
We start with the case $s\le\tfrac12$, for which the right-hand side of~\eqref{o2} is $c_s$. The upper bound follows immediately from the corresponding bound for arbitrary positive semidefinite operators, since $\md(A)\subset\pos(A)$. It remains to show that the same constant $c_s$ cannot be improved under the normalization constraint.

Consider the example from~\eqref{rhosig} with $\lambda=1$. Then $\sigma=I_d/d$ is the maximally mixed state, and the parameter in~\eqref{rr} is
$
r=k/d\leq1.
$
By varying $d$ and $k$, the attainable values of $r=k/d$ are dense in $(0,1)$. Hence, using~\eqref{rr} and the continuity of $r\mapsto r^s/(1+r)$, we obtain
\be\label{l4}
\sup_{\rho,\sigma\in\md(A)}
\frac{Q_2(\rho\|\rho+\sigma)}{Q_{1+s}(\rho\|\sigma)}
\geq
\sup_{r\in(0,1)}\frac{r^s}{1+r}\,.
\ee
Now recall that
\be
c_s=\sup_{r>0}\frac{r^s}{1+r}\,,
\ee
and that the maximizer is
$
r_s\eqdef s/(1-s).
$
Since $r_s\le1$ exactly when $s\le\tfrac12$, the right-hand side of~\eqref{l4} is equal to $c_s$. We conclude that, in the range $s\le\tfrac12$, the normalization constraint does not reduce the optimal constant.

Next we consider the case $s>\tfrac12$, for which the right-hand side of~\eqref{o2} is $\tfrac12$. The lower bound follows by taking $\sigma=\rho$. Indeed, for any $\rho\in\md(A)$,
\be
Q_2(\rho\|\rho+\rho)=\frac12\,,
\qquad
Q_{1+s}(\rho\|\rho)=1\,,
\ee
and hence
\be
\frac{Q_2(\rho\|\rho+\rho)}{Q_{1+s}(\rho\|\rho)}
=
\frac12\,.
\ee

It remains to prove the upper bound. From~\eqref{gg} we have
\be\label{q2rs}
Q_2(\rho\|\rho+\sigma)=\int_{0}^\infty \frac r{1+r}\,\d\mu(r),
\ee
where $\mu(r)$ is defined in~\eqref{mur}. In Appendix~\ref{A3} we show that, for every $r>0$ and every $s\ge\tfrac12$,
\be\label{r1}
\frac{r}{1+r}
\le
\frac12 r^{s}
+
\frac12\left(s-\frac12\right)\left(r^{-1}-1\right)\,.
\ee
Substituting this inequality into~\eqref{q2rs} and using~\eqref{leq} yields
\be
Q_2(\rho\|\rho+\sigma)
\leq
\frac12\int_{0}^\infty r^s\,\d\mu(r)
=
\frac12Q_{1+s}(\rho\|\sigma),
\ee
where the last equality follows from~\eqref{int1s}. This proves the upper bound for $s>\tfrac12$.
Combining the two cases proves~\eqref{o2}.
\end{proof}

\section{Applications}

As discussed above, Theorem~\ref{thm:1} sharpens the logarithmic trace inequality of Cheng \emph{et al.}~\cite{CGH+2025} by replacing the prefactor $c_s/s$ with the smaller Lambert-$W$-based constant $G_s$. Since this inequality underlies standard single-shot primitives such as decoupling, convex-splitting, and covering lemmas, this improvement leads directly to tighter finite-resource bounds. In this section, however, we focus on applications of Theorem~\ref{thm:3}. The main new feature of that theorem is the appearance of the sharp prefactor $c_s<1$, which improves bounds in settings where no such factor was previously available.

\subsection{Enhanced Position-Based Decoding Lemma}

The position-based decoding lemma is a powerful tool in one-shot quantum information theory. It provides a way to decode a message or subsystem by identifying its position among many candidate systems, and has become a standard ingredient in achievability proofs for one-shot quantum communication tasks. In particular, position-based decoding, often combined with convex-splitting, has been used to obtain one-shot bounds for entanglement-assisted classical and quantum communication over noisy channels, quantum network communication problems, and quantum state redistribution~\cite{AJW2019,QWW2018,AJW2018}. Here, we revisit this lemma and show that the trace inequality established in Theorem~\ref{thm:3} improves its quantitative bounds and extends its applicability to a broader class of pairwise index-symmetric states.

Let $\rho\in\md(RA)$, $\sigma\in\md(A)$, $n\in\mbb{N}$, and let
$A^n=(A_1,\ldots,A_n)$ denote $n$ copies of $A$. Consider the set of states
$\{\tau^x_{RA^n}\}_{x\in[n]}\subset\md(RA^n)$ defined by
\be\label{rcyc}
\tau^x_{RA^n}
=
\rho_{RA_x}\otimes\bigotimes_{y\in[n]\setminus\{x\}}\sigma_{A_y}\;.
\ee
The position-based decoding lemma, as stated in~\cite{AJW2019,ABS+2023}, asserts that for
\be\label{n-pbd-standard}
n\eqdef
\left\lceil
\eps\,2^{D_H^\eps(\rho_{RA}\|\rho_R\otimes\sigma_A)}
\right\rceil\;,
\ee
there exists a POVM $\{\Lambda^x_{RA^n}\}_{x\in[n]}$ such that, for all
$x\in[n]$,
\be\label{f6}
\tr\left[\Lambda^x_{RA^n}\tau^x_{RA^n}\right]\geq 1-6\eps\;.
\ee

In the next lemma, we give a variant of this result in which $Q_s$ replaces the hypothesis-testing divergence. Our lemma improves earlier bounds~\cite{BG2017,Cheng2023} by the factor $c_s$. It also applies to a broader class of states, which we call pairwise index-symmetric~\cite{Gour2025a}.

A collection of $n$ density matrices $\{\tau^x_{RA^n}\}_{x\in[n]}$ is said to be
\emph{pairwise index-symmetric} if there exist $\rho,\sigma\in\md(RA)$, with
$\rho_R=\sigma_R$, such that for every $x,y\in[n]$,
\bmyd
\be\label{cas}
\tau^x_{RA_y}
=
\begin{cases}
\rho_{RA_x} & \text{if } x=y,\\
\sigma_{RA_y} & \text{if } x\neq y.
\end{cases}
\ee
\emyd
The states in~\eqref{rcyc} are pairwise index-symmetric: they satisfy
$\tau^x_{RA_x}=\rho_{RA_x}$ for all $x\in[n]$, while for $x,y\in[n]$ with
$y\neq x$ one has $\tau^x_{RA_y}=\rho_R\otimes\sigma_{A_y}$. However, the class of pairwise index-symmetric states is larger than the product-form family in~\eqref{rcyc}. For example, one may replace the product state on the unused systems by a permutation-invariant state. Namely, let
$\omega_{A^n}$ be symmetric under permutations of the systems $A_1,\ldots,A_n$, with marginal $\omega_{A_y}=\sigma_A$ for every $y\in[n]$, and define
\be
\tau^x_{RA^n}
=
\rho_{RA_x}\otimes
\omega_{A_1\cdots A_{x-1}A_{x+1}\cdots A_n}.
\ee
These states need not have the product form~\eqref{rcyc}, but they have the same relevant two-system marginals: $\tau^x_{RA_x}=\rho_{RA_x}$ for all $x\in[n]$, and
$\tau^x_{RA_y}=\rho_R\otimes\sigma_{A_y}$ whenever $y\neq x$.

We are now ready to state the improved position-based decoding lemma. Let
$\eps\in(0,1)$, $s\in(0,1)$, and let $\rho,\sigma\in\md(RA)$ satisfy
$\rho_R=\sigma_R$. Set
\be\label{n}
n\eqdef
\left\lceil
\left(\frac{\eps}{c_s}\right)^{\frac1{1-s}}
2^{D_s(\rho_{RA}\|\sigma_{RA})}
\right\rceil\,,
\ee
and let $\{\tau^x_{RA^n}\}_{x\in[n]}\subset\md(RA^n)$ be a pairwise index-symmetric set of states satisfying~\eqref{cas}. Then:

\bmyl\label{pbdl}
There exists a POVM $\{\Lambda^x_{RA^n}\}_{x\in[n]}$ such that, for all
$x\in[n]$,
\be
\tr\left[\Lambda^x_{RA^n}\tau^x_{RA^n}\right]\geq 1-\eps\;.
\ee
\emyl

\begin{proof}
Let
\be
\eta_{RA^n}\eqdef\sum_{x\in[n]}\tau^x_{RA^n}\,.
\ee
On the support of $\eta_{RA^n}$, define
\be
\Lambda^x_{RA^n}
\eqdef
\eta_{RA^n}^{-1/2}\tau^x_{RA^n}\eta_{RA^n}^{-1/2}\;.
\ee
Then $\sum_x\Lambda^x_{RA^n}$ is the projection onto $\supp(\eta_{RA^n})$. Completing the POVM arbitrarily on the orthogonal complement does not affect the success probabilities below.

For every $x\in[n]$, we have
\ba
\tr\left[\Lambda^x_{RA^n}\tau^x_{RA^n}\right]
&=
\tQ_2\left(\tau^x_{RA^n}\big\|\eta_{RA^n}\right)\\
&\ge
Q_2\left(\tau^x_{RA^n}\big\|\eta_{RA^n}\right)\\
&\ge
Q_2\left(\tau^x_{RA_x}\big\|\eta_{RA_x}\right),
\ea
where the first inequality uses $Q_2\le \tQ_2$, and the second follows from the data-processing inequality for $Q_2$.

Combining this with~\eqref{cas}, we have
\be
\eta_{RA_x}
=
\rho_{RA_x}+(n-1)\sigma_{RA_x}.
\ee
Therefore,
\ba
\tr\left[\Lambda^x_{RA^n}\tau^x_{RA^n}\right]
&\ge
Q_2\left(\rho_{RA}\big\|\rho_{RA}+(n-1)\sigma_{RA}\right)\\
&\geq
1-c_s Q_s\left(\rho_{RA}\big\|(n-1)\sigma_{RA}\right)\\
&=
1-c_s (n-1)^{1-s}Q_s\left(\rho_{RA}\big\|\sigma_{RA}\right)\\
&\geq
1-\eps\,,
\label{subinto}
\ea
where the second inequality follows from~\eqref{lb}, and the last inequality follows from the definition of $n$ in~\eqref{n}. This completes the proof.
\end{proof}

The factor $c_s$ in~\eqref{n}, with $c_s<1$, improves on previous results by allowing a larger choice of $n$. We now apply this improvement to a single-shot classical communication task over a quantum channel.

\subsection{Classical Communication over a Quantum Channel}

We now apply Theorem~\ref{thm:3} to one-shot classical communication over a quantum channel. The point of the application is that the improved factor $c_s<1$ in the collision-divergence inequality leads directly to a sharper lower bound on the one-shot distillable communication. In the remainder of this section, we write the parameter as $\alpha\in(0,1)$, rather than $s$, in order to match the standard notation used in one-shot communication bounds. Thus
$
c_\alpha\eqdef \alpha^\alpha(1-\alpha)^{1-\alpha}
$.

We describe the task in the resource-theoretic language of channel distillation~\cite{CG2019,Gour2025}. The target resource is a noiseless classical channel of size $m$, identified with the completely dephasing channel
$\Delta_m\in\cptp(Z_A\to Z_B)$, where $Z_A$ and $Z_B$ are classical systems with
$|Z_A|=|Z_B|=m$. This channel satisfies the golden-unit relation~\cite{CG2019,Gour2025}:
\be
\Delta_m\otimes\Delta_n\cong\Delta_{mn}\qquad\forall\;m,n\in\mbb{N}.
\ee
The free transformations are local superchannels. We say that a superchannel $\Theta$ is local if, for every $\mN\in\cptp(A\to B)$, it acts as
\be
\Theta[\mN]\eqdef
\mF_{B\to B'}\circ\mN_{A\to B}\circ\mE_{A'\to A},
\ee
where $\mE$ and $\mF$ are local encoding and decoding channels, respectively.

For the achievability bound below, it is enough to consider a classical input system $X$ with $k\eqdef |X|$. Indeed, any classical communication protocol over a quantum-input channel can be viewed, after fixing the encoding, as a protocol over the induced classical-quantum channel; see, e.g.,~\cite{Wilde2013,Watrous2018}. Thus we write $\mN_{X\to B}$ for the channel and set
\be
\sigma^x_B\eqdef \mN_{X\to B}(|x\lr x|)
\qquad x\in[k]\,.
\ee 
A code of size $m$ consists of a codebook $x^m=(x_1,\ldots,x_m)\in[k]^m$ and a POVM
$\{\Lambda^z\}_{z\in[m]}$ on $B$. The optimal average-error probability, or equivalently the Choi-based conversion distance in the resource-theoretic formulation~\cite{Gour2025a}, is then
\be\label{126n}
T\left(\mN\xrightarrow{\text{\tiny LO}}\Delta_m\right)
\eqdef
\min_{x^m,\Lambda}
\frac1m\sum_{z\in[m]}
\left(1-\tr\left[\sigma^{x_z}_B\Lambda^z_B\right]\right)\,,
\ee
where the minimum is over all codebooks and all POVMs. 

Let $\eps\in(0,1)$. The $\eps$-single-shot distillable classical communication is
\be\label{a4}
\distill^\eps(\mN)
\eqdef
\max_{m\in\mbb{N}}
\left\{
\log(m)\;:\;
T\left(\mN\xrightarrow{\text{\tiny LO}}\Delta_m\right)\le\eps
\right\}.
\ee

To state the lower bound on this quantity, let $\p\in\prob(k)$ and define the cq-state
\be\label{sp}
\sigma^\p_{XB}
\eqdef
\sum_{x\in[k]}p_x|x\lr x|_X\otimes\sigma^x_B\, .
\ee
Its marginals are
\be
\sigma^\p_X\eqdef\sum_{x\in[k]}p_x|x\lr x|_X\, ,
\qquad
\sigma^\p_B\eqdef\sum_{x\in[k]}p_x\sigma^x_B\, .
\ee
For $\alpha\in(0,1)$, define the R\'enyi mutual information of the channel by
\be
I_\alpha(X:B)_\mN
\eqdef
\max_{\p\in\prob(k)}
D_\alpha\left(\sigma^\p_{XB}\middle\|\sigma^\p_X\otimes\sigma^\p_B\right).
\ee
Then the improved collision-divergence inequality gives the following lower bound, whose additive improvement is governed by the binary entropy 
\be
h(\alpha)\eqdef-\alpha\log(\alpha)-(1-\alpha)\log(1-\alpha)\,.
\ee

\bmyt\label{thmlub}
For every $\alpha\in(0,1)$,
\ba\label{151}
&\distill^\eps(\mN)\\
&\geq
I_{\alpha}(X:B)_\mN
-
\frac{1}{1-\alpha}\log\frac1\eps
+
\frac{h(\alpha)}{1-\alpha}\,.
\ea
\emyt
\begin{remark}
Eq.~(12) of~\cite{Cheng2023} gives, for every $\delta\in[0,\eps)$ and every $\p\in\prob(k)$,
\be
\distill^\eps(\mN)
\geq
D_H^\delta\left(\sigma^\p_{XB}\middle\|\sigma^\p_X\otimes\sigma^\p_B\right)
+
\log(\eps-\delta).
\ee
Combining this with the universal bound~\cite{AMV2012,Gour2026}
\be
D_H^\delta(\rho\|\sigma)
\geq
D_\alpha(\rho\|\sigma)
+
\log\frac1{1-\alpha}
-
\frac{\alpha}{1-\alpha}\log\left(\frac{\alpha}{\delta}\right)
\ee
yields an $I_\alpha$-bound with additive term
\be
f_\alpha(\eps,\delta)
\eqdef
\log\frac1{1-\alpha}
-
\frac{\alpha}{1-\alpha}\log\left(\frac{\alpha}{\delta}\right)
+
\log(\eps-\delta).
\ee
Optimizing over $\delta$ gives
\be
\sup_{\delta\in[0,\eps)} f_\alpha(\eps,\delta)
=
\frac{1}{1-\alpha}\log(\eps)\,,
\ee
with the optimum attained at $\delta=\alpha\eps$. Thus the indirect route through the hypothesis-testing divergence recovers only the usual negative finite-error correction
$-(1-\alpha)^{-1}\log(1/\eps)$.

The point of Theorem~\ref{thmlub} is that Theorem~\ref{thm:3} allows one to work directly with the collision-divergence bound and convert it into a $Q_\alpha$-based bound with the sharp factor $c_\alpha$. Equivalently, since
\be
\log\frac1{c_\alpha}
=
h(\alpha),
\ee
this factor contributes the positive additive term $h(\alpha)/(1-\alpha)$ in~\eqref{151}. Hence the bound in~\eqref{151} contains the usual finite-error penalty
$-(1-\alpha)^{-1}\log(1/\eps)$, but also a positive entropy correction coming from the optimal trace inequality. In particular, the total correction is positive precisely when $\eps>c_\alpha$.
\end{remark}

\begin{proof}
For every codebook $x^m\in[k]^m$, set
\be
\eta_B^{x^m}\eqdef\sum_{z\in[m]}\sigma^{x_z}_{\!B}\,.
\ee
Consider the square-root measurement
\be
\Lambda^z_B\eqdef
\left(\eta^{x^m}_B\right)^{-1/2}\sigma^{x_z}_{\!B}\left(\eta^{x^m}_B\right)^{-1/2},
\qquad z\in[m]\,.
\ee
Although $\Lambda^z_B$ depends on the codebook $x^m$, we suppress this dependence in the notation for simplicity.
Substituting this POVM into~\eqref{126n} gives
\be\label{start}
T\left(\mN\xrightarrow{\text{\tiny LO}}\Delta_m\right)
\leq
1-\max_{x^m\in[k]^m}
\frac1m\sum_{z\in[m]}
\tQ_2\left(\sigma^{x_z}_{\!B}\big\|\eta_B^{x^m}\right).
\ee

We lower bound the maximal average success probability by averaging over random codebooks. Fix $\p\in\prob(k)$ and draw
$x^m=(x_1,\ldots,x_m)$ i.i.d.\ according to
$p_{x^m}\eqdef p_{x_1}\cdots p_{x_m}$. Then
\ba\label{io}
T&\left(\mN\xrightarrow{\text{\tiny LO}}\Delta_m\right)\\
&\leq
1-
\frac1m\sum_{z\in[m]}\sum_{x^m\in[k]^m}
p_{x^m}
\tQ_2\left(\sigma^{x_z}_{\!B}\big\|\eta_B^{x^m}\right)\,.
\ea

For each $z\in[m]$, define
\be
\tau^z_{X^mB}
\eqdef
\sum_{x^m\in[k]^m}
p_{x^m}|x^m\lr x^m|_{X^m}\otimes\sigma^{x_z}_B,
\ee
and
\ba
\tau_{X^mB}
&\eqdef
\sum_{x^m\in[k]^m}
p_{x^m}|x^m\lr x^m|_{X^m}\otimes\eta_B^{x^m}\\
&=
\sum_{z\in[m]}\tau^z_{X^mB}\,.
\ea
By the direct-sum property of $\tQ_2$, the DPI, and the fact that $\tQ_2\ge Q_2$ we get
\ba\label{io2}
\sum_{x^m\in[k]^m}
p_{x^m}
\tQ_2\left(\sigma^{x_z}_{\!B}\big\|\eta_B^{x^m}\right)
&=
\tQ_2\left(\tau^z_{X^mB}\big\|\tau_{X^mB}\right)\\
&\geq
\tQ_2\left(\tau^z_{X_zB}\big\|\tau_{X_zB}\right)\\
&\geq Q_2\left(\tau^z_{X_zB}\big\|\tau_{X_zB}\right)\,.
\ea
Tracing out all systems except $X_zB$, and relabeling $X_z$ as $X$, gives
\be
\tau^z_{X_zB}\equiv\sigma^\p_{XB}\,,
\qquad
\tau_{X_zB}
\equiv
\sigma^\p_{XB}
+
(m-1)\sigma^\p_X\otimes\sigma^\p_B\,.
\ee
Therefore, by the lower bound in Theorem~\ref{thm:3},
\ba\label{io3}
Q_2\big(&\tau^z_{X_zB}\big\|\tau_{X_zB}\big)\\
&=
Q_2\left(
\sigma^\p_{XB}
\middle\|
\sigma^\p_{XB}
+
(m-1)\sigma^\p_X\otimes\sigma^\p_B
\right)\\
&\geq
1-
c_\alpha (m-1)^{1-\alpha}
Q_\alpha\left(
\sigma^\p_{XB}
\middle\|
\sigma^\p_X\otimes\sigma^\p_B
\right).
\ea
Combining~\eqref{io}, \eqref{io2}, and~\eqref{io3}, we obtain, for every
$\p\in\prob(k)$,
\be\label{ub2}
T\left(\mN\xrightarrow{\text{\tiny LO}}\Delta_m\right)
\leq
c_\alpha (m-1)^{1-\alpha}
Q_\alpha\left(
\sigma^\p_{XB}
\middle\|
\sigma^\p_X\otimes\sigma^\p_B
\right).
\ee
Choosing $\p$ such that
\be
Q_\alpha\left(
\sigma^\p_{XB}
\middle\|
\sigma^\p_X\otimes\sigma^\p_B
\right)
=
2^{(\alpha-1)I_\alpha(X:B)_\mN}\,,
\ee
we get that every integer $m$ satisfying
\be
c_\alpha (m-1)^{1-\alpha}
2^{(\alpha-1)I_\alpha(X:B)_\mN}
\leq
\eps
\ee
is achievable. Taking
\be
m=
1+
\left\lfloor
\left(\frac{\eps}{c_\alpha}\right)^{\frac1{1-\alpha}}
2^{I_\alpha(X:B)_\mN}
\right\rfloor\,,
\ee
and using $1+\lfloor x\rfloor\ge x$ for $x>0$, we obtain
\be
\log(m)
\geq
I_\alpha(X:B)_\mN
-
\frac1{1-\alpha}\log\frac1\eps
+
\frac{h(\alpha)}{1-\alpha}\,,
\ee
where we used $\log(1/c_\alpha)=h(\alpha)$. This proves~\eqref{151}.
\end{proof}

\section{Conclusions}

We established sharp quantum trace inequalities that improve finite-resource bounds in single-shot quantum information theory. The first main result is an optimal logarithmic trace inequality, replacing the prefactor $c_s/s$ in the bound of Cheng \emph{et al.}~\cite{CGH+2025} by the strictly smaller constant $G_s$. This constant is determined by the optimal scalar inequality $\log(1+r)\le G_s r^s$ and admits a closed-form expression in terms of the Lambert $W$ function. The improvement is particularly significant in the relative-entropy limit $s\to0$, where the ratio between the previous prefactor and the optimal one tends to $e$.

We proved that $G_s$ is the optimal universal constant for arbitrary positive operators: no smaller constant can bound $Q(\rho\|\sigma)$ by $Q_{1+s}(\rho\|\sigma)$ uniformly in finite dimensions. This optimality is already witnessed by commuting examples and remains unchanged if $Q_{1+s}$ is replaced by the larger sandwiched quantity $\tQ_{1+s}$. We also resolved the normalized-state version. For density matrices, the optimal constant exhibits a sharp threshold: it is $G_s$ for $s\le 1/(2\log(2))$, while for $s>1/(2\log(2))$ it drops to $\log(2)$. Thus the fully quantum normalized problem is completely characterized.

A second set of results concerned the collision-type trace inequalities in Theorem~\ref{thm:3}. We showed that the constant $c_s=s^s(1-s)^{1-s}$ is optimal in both inequalities for arbitrary positive operators: the same commuting examples attain equality, and no smaller constant works uniformly. Although $c_s$ is not optimal for the logarithmic inequality, it is the correct constant for these collision-divergence inequalities. Moreover, the upper collision bound implies the earlier logarithmic estimate with prefactor $c_s/s$, showing that it is a stronger structural statement.

For normalized states, the two collision inequalities behave differently. For the lower collision bound, the same constant $c_s$ remains optimal for density matrices for all $s\in(0,1)$, so no threshold phenomenon occurs. By contrast, the upper collision bound does exhibit a threshold: the sharp constant is $c_s$ for $s\le1/2$, while for $s>1/2$ it drops to $1/2$. Thus normalization affects the two collision inequalities in distinct ways, both of which can be characterized exactly.

The proofs are based on an iterative Riemann--Stieltjes integration-by-parts method within the operator layer-cake framework. This method lifts sharp scalar inequalities to the noncommutative setting without loss, by integrating against the positive measure generated by the threshold projections $\{\rho>r\sigma\}$ and then integrating by parts again to recover the corresponding layer-cake R\'enyi quantity. In this way, the layer-cake representation becomes not only a representation formula, but also a mechanism for preserving optimal constants at the operator level.

We also showed that these sharper trace inequalities lead to improved single-shot information-theoretic bounds. In particular, the lower collision inequality yields an enhanced position-based decoding lemma with the improved factor $c_\alpha<1$, and applies beyond the standard product construction to pairwise index-symmetric families of states. We then used the same bound to obtain a tightened one-shot lower bound on classical communication over a quantum channel, formulated as the distillation of a noiseless classical channel under local superchannels. In this bound, the factor $c_\alpha$ appears as the positive additive contribution $h(\alpha)/(1-\alpha)$, which can partially offset the usual negative finite-error correction and can make the total correction positive when $\eps>c_\alpha$. These applications illustrate how improvements in trace inequalities translate directly into sharper finite-resource performance bounds.

More broadly, the results identify exact optimality barriers for a class of layer-cake and R\'enyi-divergence methods in single-shot quantum information. The constants obtained here are not artifacts of the proof, but the best possible constants within the inequalities considered, both for arbitrary positive operators and for normalized quantum states. An interesting direction for future work is to understand when the layer-cake or sandwiched R\'enyi quantities appearing in these bounds can be replaced by measured R\'enyi divergences. Since measured divergences are smaller, such replacements could in principle lead to sharper exponents, but known classical-to-quantum lifting methods introduce pinching-dependent prefactors. Determining when this trade-off is favorable remains an important question.

\section*{Acknowledgment}
This research was supported by the Israel Science Foundation under Grant No. 1192/24.

\bibliographystyle{apsrev4-2}
\bibliography{QRT} 

\newpage

\onecolumngrid

\appendix

\begin{center}
{\large\bfseries Appendix}
\end{center}
%\vspace{1em}

\section{A Scalar Maximization Lemma for Normalized States}\label{A1}

Consider the function
\be
g_s(r)\eqdef \frac{\log(1+r)}{r^s}\,, \qquad r>0\,,
\ee
so that $G_s=\sup_{r>0} g_s(r)$.

\bmyl
Let $s_0\eqdef 1/(2\log(2))$. For every $s\in(0,1)$, the function
$g_s$ has a unique maximizer $r_s\in(0,\infty)$. Moreover,
\be
r_s>1 \quad\text{if } s<s_0\,,\qquad
r_s=1 \quad\text{if } s=s_0\,,\qquad
r_s<1 \quad\text{if } s>s_0\, .
\ee
\emyl

\begin{proof}
 To locate the maximizer, we analyze $g_s$.
A direct computation yields
\be
g_s'(r)=\frac{1}{r^{s+1}}\left(\frac{r}{1+r}-s\log(1+r)\right)\,.
\ee
Thus, the sign of $g_s'(r)$ is determined by
\be
h_s(r)\eqdef \frac{r}{1+r}-s\log(1+r)\,.
\ee
Differentiating,
\be
h_s'(r)=\frac{1-s(1+r)}{(1+r)^2}\,.
\ee
Hence $h_s'$ vanishes exactly once, at $r=(1-s)/s$, and therefore $h_s$ is strictly increasing on $(0,(1-s)/s)$ and strictly decreasing on $((1-s)/s,\infty)$. In particular, $h_s$ has a unique zero, and thus $g_s$ admits a unique maximizer $r_s$, characterized by $h_s(r_s)=0$.

Evaluating $h_s$ at $r=1$ gives
\be
h_s(1)=\frac12 - s\log(2).
\ee
Thus, $h_s(1)>0$ for $s<s_0$, $h_s(1)=0$ for $s=s_0$, and $h_s(1)<0$ for $s>s_0$. Since $h_s$ has a unique zero and is increasing before its maximizer and decreasing afterward, it follows that this zero lies to the right of $1$ when $s<s_0$, at $1$ when $s=s_0$, and to the left of $1$ when $s>s_0$. 
Consequently, the maximizer $r_s$ of $g_s$ satisfies
$r_s > 1$ if $s<s_0$, $r_s = 1$ if $s=s_0$, and $r_s < 1$  if $s>s_0$.
\end{proof}

\section{Proof of the Inequality~\eqref{auxineq}}\label{A2}

Let $s_0\eqdef 1/(2\log(2))$. 
\bmyl
For every $s\ge s_0$ and every $r>0$,
\be
\frac{\log(1+r)}{\log(2)}
\le
r^s+(s-s_0)(r^{-1}-1)\,.
\ee
\emyl

\begin{proof}
For fixed $s\ge s_0$, define
\be
F_s(r)\eqdef
r^s+(s-s_0)(r^{-1}-1)
-\frac{\log(1+r)}{\log(2)} .
\ee
It is enough to prove that $F_s(r)\ge0$ for all $r>0$.

First consider $s=s_0$. By Appendix~\ref{A1}, the function
$r\mapsto \log(1+r)/r^{s_0}$ attains its maximum at $r=1$. Hence
\be
\frac{\log(1+r)}{\log(2)}
\le r^{s_0},
\qquad r>0,
\ee
and therefore $F_{s_0}(r)\ge0$.

It remains to show that $F_s(r)$ is nondecreasing in $s$. For fixed $r>0$,
\be
\partial_s F_s(r)
=
r^s\log(r) + r^{-1}-1.
\ee
Multiplying by $r>0$, we obtain
\be
r\,\partial_sF_s(r)
=
r^{s+1}\log(r)-(r-1).
\ee
If $r\ge1$, then $r^{s+1}\log(r)\ge r\log(r)$. If $0<r\le1$, then
$r^{s+1}\le r$ and $\log(r)\le0$, so again
$r^{s+1}\log(r)\ge r\log(r)$. Therefore, for all $r>0$,
\be
r^{s+1}\log(r)\ge r\log(r)\ge r-1,
\ee
where the last inequality is the standard bound $r\log(r)\ge r-1$.
Thus $\partial_sF_s(r)\ge0$, and hence
\be
F_s(r)\ge F_{s_0}(r)\ge0.
\ee
This proves~\eqref{auxineq}.
\end{proof}

\section{Proof of the Inequality~\eqref{r1}}\label{A3}

\bmyl
For every $s\ge \tfrac12$ and every $r>0$,
\be
\frac{r}{1+r}
\le
\frac12 r^s
+
\frac12\left(s-\frac12\right)(r^{-1}-1).
\ee
\emyl

\begin{proof}
For fixed $s\ge\tfrac12$, define
\be
F_s(r)\eqdef
\frac12 r^s
+
\frac12\left(s-\frac12\right)(r^{-1}-1)
-
\frac{r}{1+r}.
\ee
It is enough to prove that $F_s(r)\ge0$ for all $r>0$.

First consider $s=\tfrac12$. Then
\be
F_{1/2}(r)
=
\frac12 r^{1/2}
-
\frac{r}{1+r}
=
\frac{r^{1/2}}{2(1+r)}(1+r-2\sqrt r)
\ge0,
\ee
where the last inequality follows from $1+r\ge2\sqrt r$.

It remains to show that $F_s(r)$ is nondecreasing in $s$. For fixed $r>0$,
\be
\partial_s F_s(r)
=
\frac12 r^s\log r
+
\frac12(r^{-1}-1).
\ee
Multiplying by $2r>0$, we get
\be
2r\,\partial_s F_s(r)
=
r^{s+1}\log r-(r-1).
\ee
If $r\ge1$, then $r^{s+1}\log r\ge r\log r$. If $0<r\le1$, then
$r^{s+1}\le r$ and $\log r\le0$, so again
$r^{s+1}\log r\ge r\log r$. Hence, for all $r>0$,
\be
r^{s+1}\log r\ge r\log r\ge r-1,
\ee
where the last inequality is the standard bound $r\log r\ge r-1$.
Therefore $\partial_sF_s(r)\ge0$, and so
\be
F_s(r)\ge F_{1/2}(r)\ge0.
\ee
This proves~\eqref{r1}.
\end{proof}

 \end{document}